# SELECTING 3D CHAOTIC FLOW STATES FOR ACCELERATED DNA REPLICATION IN MICRO-SCALE CONVECTIVE PCR

**Aashish Priye, Radha Muddu, Yassin A. Hassan, and Victor M. Ugaz**
*Department of Chemical Engineering, Texas A&M University, College Station, TX 77843, USA*

**ABSTRACT**

Micro-scale flow in cylindrical geometries can harness chaotic advection to perform complex thermally activated biochemical reactions such as the polymerase chain reaction (PCR). We have applied a 3D computational fluid dynamics model to resolve the complex flow patterns in such geometries. The resulting 3D flow trajectories are then used as input to a kinetic model to resolve the time evolution of DNA replication process. A simple mass action kinetic model was developed to couple these biochemical reactions with the intricate flow. Residence time analysis of virtual particles in the flow revealed that the flow has a strong chaotic component in wider geometries in comparison with taller geometries (quasi periodic motion). This work shows, for the first time that the chaotic aspect of the flow field plays a key role in determining the strength of the coupling between the reactions and the flow. Our model can quantify the doubling times of these reactions capturing the lag, exponential and plateau phases of PCR. It predicts that doubling times are lower in wider geometries, in agreement with experimental results.

## INTRODUCTION

PCR has been performed in a variety of microfluidic devices, mostly with the aim of reducing the reaction time. Rayleigh-Bénard convection offers potential to enable rapid PCR and possesses an added advantage of being portable[1-3]. It is based on the simple idea of natural convection where the PCR reagent mixture is confined in cylindrical vessel. The temperature of the bottom surface is maintained at a higher temperature than the top. This causes a buoyancy driven instability followed by a convective flow inside the geometry. The onset of flow and transition to convective turbulence is determined by a dimensionless number known as the Rayleigh number ($Ra = [g\beta(T_2-T_1)h^3]/\nu\alpha$ ; where $\beta$ is the fluid's thermal expansion coefficient, $g$ is gravitational acceleration, $T_1$ and $T_2$ are the temperatures of the top (cool) and bottom (hot) surfaces respectively, $h$ is the height of the fluid layer, $\alpha$ is the thermal diffusivity, and $\nu$ is the kinematic viscosity). The reagents are cycled continuously through different temperature zones where repeated denaturing (~95 °C), annealing (50-60 °C) and extension (72 °C) reactions results in a chain reaction, exponentially increasing the number of copies of the target DNA sequence (Fig 1).

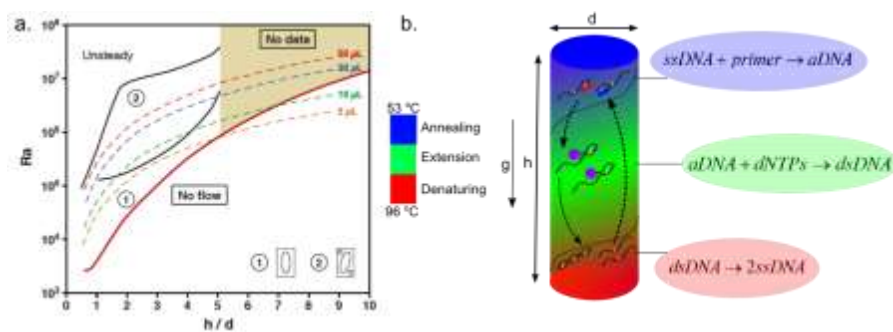

*Figure 1 PCR in cylindrical geometries (a)Micro-scale thermal convection generates a multiplicity of flow regimes at h/d>1(b)Denaturing, Annealing and extension reactions at different zones in the cylindrical reactor.*

## THEORY

Although the convective reactor configuration seems straightforward, the resulting micro-scale flows are surprisingly complex. This is due to the fact that the flow is quite sensitive to geometrical and thermal parameters and tuning these parameters desirably results in the onset of chaotic flow and convective turbulence. Under such condition the fluid elements no longer shows periodicity in flow through the volume but enters a regime of quasi periodic motion followed by full blown chaos. For $h/d = 3$ the flow trajectories changes from 2D circulatory with quasi periodic flow (KAM boundaries on the Poincaré maps) to a 3D chaotic flow when the temperature gradient is changed from 5 °C to 35 °C. When its further increased to 100 °C the flow field becomes even more disordered with the emergence of multiple convection loops. The chaotic component of the flow can be quantified with the correlation dimension parameter which is obtained from time evolution of temperature of an individual fluid element in the flow.

It now becomes important to quantify the implications of chaotic flow to thermally activated biochemical reactions such as the PCR. The three reactions of PCR are localized within three spatially distinct zones in the reactor. The volumes of these zones are bounded by upper and lower temperatures corresponding to each reaction. A Lagrangian representation of the local temperature around the fluid element was tracked as a function of time from which the residence times and the frequencies of entry within a particular reaction zone were determined for different flow fields. To globally quantify DNA amplification within the entire reactor volume, chemical reaction kinetics and species transport was incorporated in the 3D computational flow model (Fig. 1b). Doubling times for the DNA replication can be extracted from the kinetic data. These doubling times represent global influence of the chaotic strength of flows in different geometries.

**EXPERIMENTAL/SIMULATION**

Cylinders of diameter = 1.5 mm and varying heights were created and meshed in gambit and grid independence study was performed initially. The 3D steady state Navier-Stokes equations and the energy equations were solved with the boussinesq approximation in consideration of the buoyancy driven forces to obtain the flow field for the prescribed boundary conditions (bottom = 96 °C and top = 53 °C). The fluid in the simulation was taken as water and all properties were determined at 75 °C. To investigate the effect of the chaotic flow, individual stream traces were examined for two different geometries: a taller cylinder with a $h/d = 9$ and a shorter but wider cylinder with $h/d = 3$. The time a fluid element spends in a particular zone (residence time) and the number of times that fluid elements passes through the zone were determined. This was then done for 300 randomly chosen stream traces. A distribution of residence times and frequencies of ingress in the zones were obtained for each reaction zone and visualized as histograms in (Fig. 2c). For the taller geometry ($h/d = 9$), individual fluid elements on an average spent more time in each zone whereas the number of times a fluid elements passes through each zone is more for the wider geometry ($h/d = 3$). This can be attributed to the chaotic nature of flow in the $h/d = 3$ case in comparison to the quasi periodic motion in the $h/d = 9$ case.

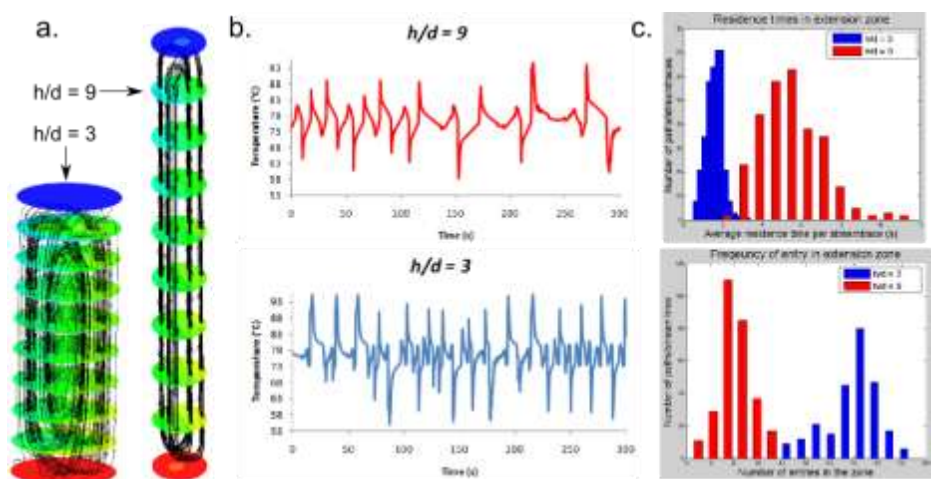

*Figure 2 (a) Individual stream traces for aspect ratios 3 and 9. Quasi periodic flow in h/d = 9 where the reagents are circulated along closed pseudo 2D paths, chaotic flow in h/d = 3 where the fluid elements follow complex 3D trajectories spanning the entire volume. (b) Thermal profiles of an individual fluid element for both h/d = 3 and h/d = 9 case .(c ) For h/d =9 the average time spent in the extension zone on an average is more than h/d = 3. However the number of times the reagents pass through the extension zone on an average is more for h/d = 3.*

*Table 1. Mass action kinetics of polymerase chain reaction*

| Reactions | Molecularity | Temperature range | Rate constants | Steps |
|---|---|---|---|---|
| dsDNA→2ssDNA | Unimolecular | 92-97 °C | 10/sec | Denaturing |
| 2ssDNA→dsDNA | Bimolecular | 55-60 °C | 106 /M.s | |
| ssDNA+primer→aDNA | Bimolecular | 55-60 °C | 5 x 106 /M.s | Annealing |
| aDNA→ssDNA+primer | Unimolecular | 92 - 97 °C | 100/sec | |
| aDNA+dNTPs→dsDNA | Bimolecular | 70-74 °C | 106 /M.s | Extension |

We next applied a kinetic model to better understand the implication of this coupling between the flow and reaction. Mass action kinetics were assumed for the denaturing, annealing and extension reactions[4]. The annealing and extension reactions were modeled as bimolecular reactions to capture the kinetics of the primers and the dNTPs also. Reversibility of denaturing and annealing reactions were also taken into account. The rate constants were taken as bell shaped Gaussian distributions whose maximum values are given in table 1.These distributions localized each step of the reaction within its respective temperature zones. The initial concentrations were determined from the initial concentrations of the reagents used in the lab to do PCR. The transient species transport equation for each species was then solved coupled with flow and energy equations to obtain a time evolution of concentration for each species. The bio-molecules were assigned the properties of water medium.

**RESULTS AND DISCUSSION**

The model does well to capture the important growth stages i.e the lag, exponential and plateau phases of DNA amplification. Initially there are no ssDNA and aDNA, therefore only denaturing reaction occurs which results in the depletion in the dsDNA copies (lag phase). However when adequate amount of ssDNA and aDNA are produced in the system, the amplification of the dsDNA takes exponential form. Finally when the primers are exhausted through the annealing reaction, the exponential increase is hindered and the plateau phase is observed (Fig 3a). It is the exponential phase which of interest to us from which a doubling time (time for the number of DNA copies to double) can be extracted as a parameter. This can be regarded equivalent to a cycle time in the conventional thermo-cycler. Based on this formulation the doubling times for different aspect ratios reveals that the chaotic interplay in the flow enhances PCR in wider geometries giving lower doubling times in quantitative agreement with experimental results[3].

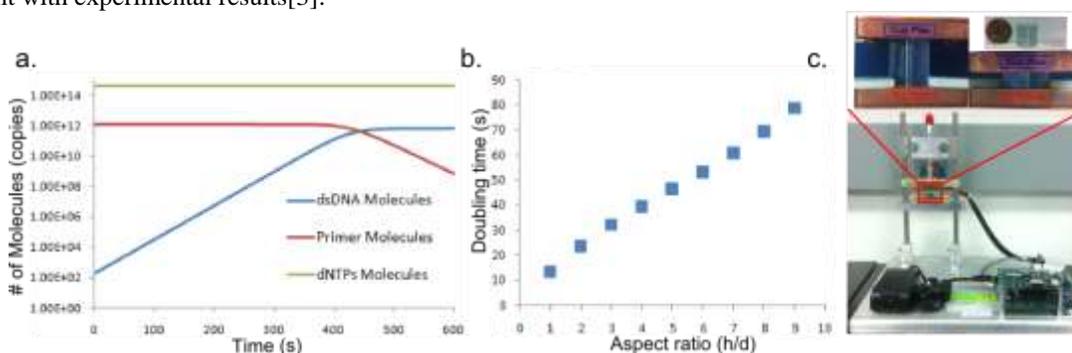

*Figure 3(a) Number of dsDNA molecules increases exponentially until primers are exhausted whereas the number of dNTP molecules remains almost unchanged. (b) Doubling time increases for geometries with higher aspect ratios (h/d). (c) The experimental setup*

**CONCLUSION**

The trajectories of individual fluid elements reveal that even thought the reagents in the taller geometry ($h/d = 9$) spend more time in the reaction zones in comparison with the wider geometry ($h/d =3$), the chaotic flow field in the wider geometry ensures that the reagents span the entire volume and enters the reaction zone more number of times. This was further verified by coupling a kinetic model with the flow to obtain rate at which the dsDNA is formed. It was found that the doubling time was lower for the wider cylinder. This verifies that quantitatively that the chaotic aspect of the flow in cylinders of lower aspect ratios enhances the yield of thermally activated biochemical reactions such as the PCR.

**ACKNOWLEDGEMENTS**

Nice words and appreciation to someone for something they've done to assist you with your paper.